\documentstyle[epsf,11pt]{article}

\begin{document}
\baselineskip=5mm

\vspace*{0.1cm}

\begin{center}

{\huge \bf Modulation of Harmonic Emission Spectra \\*[2mm]
           from~Intense Laser-Plasma Interactions} \\*[4mm]

\vspace*{9mm}

{\Large T.J.M.\ Boyd$^{1}$ and R.\ Ondarza-Rovira$^{2}$ }

\

\noindent \hspace*{-.5cm}$^{1}${\it Centre~for~Physics,~University~of~Essex,~Wivenhoe~Park,~Colchester~CO4~3SQ,~U.K.}

\vspace*{1mm}

\noindent \hspace*{-.4cm}$^{2}${\it ININ, A.P.\ 18-1027, 
M\'{e}xico 11801, D.F., Mexico}

\end{center} 
\vspace*{7mm}

\begin{center}
\parbox{14cm}{\small {\bf Abstract.} 
\noindent We report results from a series of PIC simulations 
carried out to
investigate the interaction of intense laser light with overdense 
plasma and in particular, aspects of the harmonic emission spectrum
generated in these interactions. Recent observations of emission in 
this range by Watts {\it et al.\ }[1] show emission above the 
20th 
harmonic extending to the 37th with conversion efficiencies of order
$10^{-6}$ with this spectrum showing distinct, if variable, 
modulation of between two and four harmonics.
This spectral feature proved highly sensitive to the incident 
intensity, with more pronounced modulation at higher intensities. 
These authors also reported results from a set of model PIC 
simulations which showed a modulated emission spectrum across the 
entire harmonic range, but with a wider modulation bandwidth. 

\

\noindent In the results to be presented we report on features of 
the emission
spectrum across a wide parameter range and its dependence on the 
detailed structure of the plasma density profile. While we do see 
evidence of a modulated spectrum, this appears to be sensitive not 
only to the incident intensity but to the plasma density profile. 
In addition plasma emission present further complicates the 
harmonic emission spectrum. Our results lead us to conclude that 
attempts to interpret modulation in emission spectra in terms of one
particular theoretical model and from this to infer the dynamics of
the critical surface may be misguided. }

\end{center}

\

\begin{center}
\renewcommand{\thesection}{I.}
\section{INTRODUCTION}
\end{center}

\vspace*{0.5cm}

\noindent The interaction of high power laser pulses with ultradense
plasma has proved a rich source of new plasma phenomena as varied as
channel formation and the generation of intense-localized magnetic
fields. One of the earliest laser-plasma interactions to be studied
in detail - and one of enduring interest - is the generation of
multiple harmonics of the incident light. Different mechanisms for
harmonic generation apply across differing intensity ranges and target
characteristics but perhaps most attention now attaches to harmonic
production by short intense laser pulses incident on solid targets.
Short pulses without significant pre-pulse create target plasmas 
free from extensive underdense coronal regions.

\

\noindent Harmonic generation is not only of intrinsic interest but 
holds promise as a diagnostic of plasma conditions in the surface
interaction region and, potentially of other physical processes
localized there. As well as targets of dense slab plasmas the other
determinant for efficient harmonic production is the brightness of the
source. For light of intensity $I$ and wavelength $\lambda_L$, values
of $I\, \lambda_{L}^{2}$ in the range $10^{18}-10^{19}$ W cm$^{-2}$ 
$\mu$m$^2$ typically generate a harmonic range of 50 harmonics and
Norreys {\it et al.\ }[2] detected harmonics up to the 75th (with
conversion efficiency of $10^{-6}$) for 1.053 $\mu$m wavelength light
interacting with a solid target. Such a harmonic range holds potential
as a source of coherent XUV radiation. The harmonics in the spectrum 
of reflected light are generated by electrons accelerated across the
vacuum-solid interface by the electric field of the pulse.

\

\noindent The spectral characteristics of the reflected light are 
broadly understood. Conditions favouring optimal harmonic 
generation include intensities in the range $10^{18}-10^{19}$ 
W cm$^{-2}$ $\mu$m$^2$ with plasma densities of a few tens $n_c$, where
$n_c$ denotes the critical density. Typically this combination of
intensity and target plasma density gives rise to a spectrum of 50
or more harmonics with a conversion efficiency of $10^{-6}$.
An important transition takes place at an intensity threshold below
which plasma expansion occurs thus creating a region of underdense
plasma and above which channel boring and ponderomotive steepening are
dominant.

\

\noindent Particle-in-Cell (PIC) simulations have been invaluable
in understanding and characterizing this harmonic emission. Gibbon
[3] used this approach to good effect in identifying the physical
mechanism underlying harmonic generation, attributing it to those
electrons that make vacuum excursions and then free-stream back into
the target plasma. Gibbon's work showed that the current sources
generating the harmonics are localized just within the overdense
plasma boundary. These simulations led to an empirical expression for
the efficiency of high-order harmonic generation 
$\eta_m \sim 9 \times 10^{-5} {\left( I_{18} \lambda_L^2 \right)}^2
\, {\left( 10/m \right)}^{-5}$ where $I_{18}$ denotes light intensity
in units of $10^{18}$ W cm$^{-2}$.

\

\noindent In the work presented here we use a PIC code to further
interpret aspects of the reflected light spectrum and to extend these
simulations over a wider range of parameters. Boyd and Ondarza 
[4,5] examined in detail the effects of plasma line emission
on the harmonic spectrum.
Strong plasma emission was recorded, with harmonics up to the fifth 
detected in both the reflected and transmitted spectra. A robust 
feature of the reflected spectra appeared on the blue side of the 
plasma line between (1.2-1.6) $\omega_p$, where $\omega_p$ is the
plasma frequency. In this work jets of electrons of relativistic
energy penetrating the ultra-dense plasma generated strongly 
nonthermal Langmuir waves. The spectrum of Langmuir waves was 
localized at the front edge of the plasma slab. This localization is
important for the subsequent coupling of the plasma waves to the 
radiation field by means of the steep density gradient in the 
(perturbed) peak density region. Fig.\ 1 reproduces the spectrum
obtained by Boyd and Ondarza [5] showing the reflected harmonic
spectrum and the plasma line for $n_e/n_c=30$ and $a_0=0.5$. In this
case the feature between $\omega_p$ and $2\, \omega_p$ shows up as
enhanced harmonic lines between $8\, \omega_L$ and $12\, \omega_L$,
in contrast to the broad less well-resolved feature recorded for 
other choices of parameters.

\renewcommand{\thefigure}{1}
\begin{figure}[h]  
\centering   \leavevmode
{\bf \epsfxsize=13.0cm \epsfysize=6.0cm \epsffile{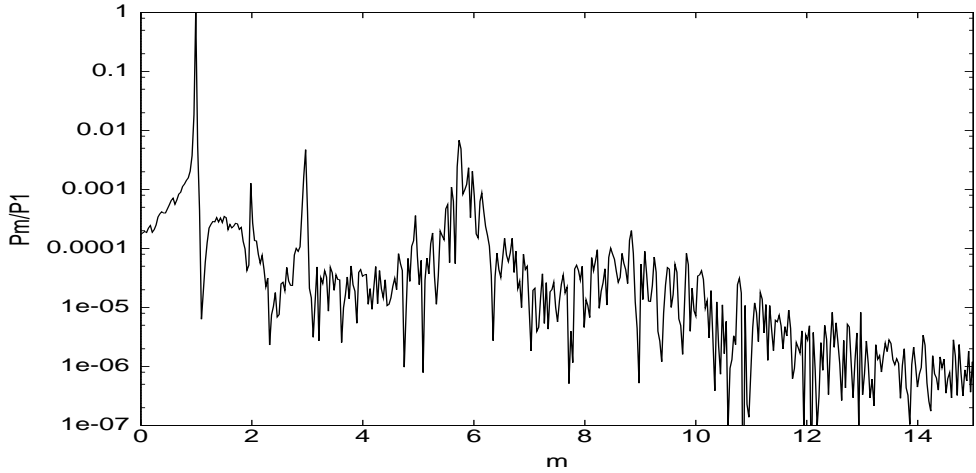} }
\caption{\small \it Reflected harmonic spectrum for $n_e/n_c = 30$
and $a_0=0.5$.}
\end{figure}
\

\noindent We were prompted to re-examine the plasma line effects on 
the spectrum by results reported by Watts {\it et al.\ }[1].
In this work using light intensities up to $1.0 \times 10^{19}$ W
cm$^{-2}$ incident on a solid surface, harmonics up to the 37th were
recorded with a conversion efficiency $\geq 10^{-7}$. The striking 
feature of the spectra at the highest intensity is a distinct if
irregular modulation with a frequency between 2 and 4 times that
of the incident light (Fig.\ 2). Watts {\it et al.\ }found that 
although 
modulations peak at different harmonics they were observable across
the entire spectral range although their published spectra show only a 
part of this range, and that at relatively low conversion efficiencies.

\renewcommand{\thefigure}{2}
\begin{figure}[h]  
\centering   \leavevmode
{\bf \epsfxsize=13.0cm \epsfysize=7.0cm \epsffile{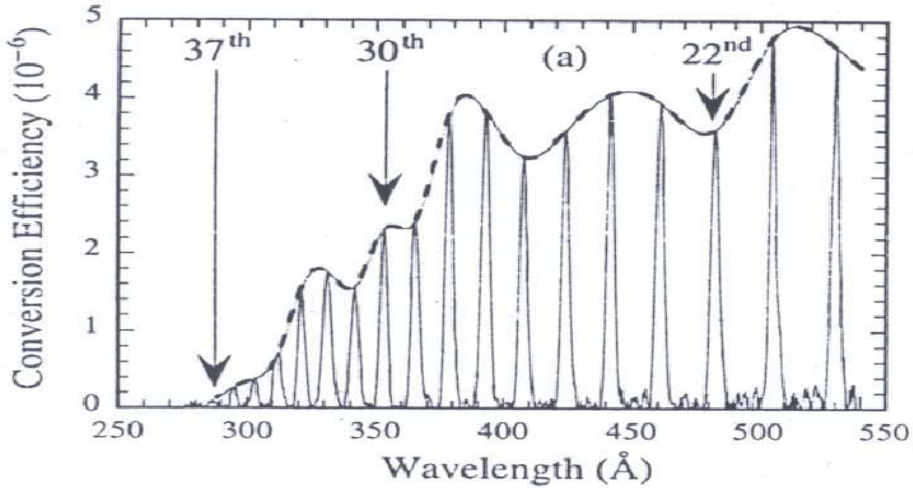} }
\caption{\small \it Modulation of harmonic spectra, for a
laser-dense plasma interaction, $I\, \lambda_L^2 \sim 10^{19}$ W
cm$^{-2}$ $\mu$m$^2$, from Watts et al.\ [1].}
\end{figure}

\noindent The corresponding spectrum at $I \lambda_L^2 \sim 10^{18}$ 
W cm$^{-2}$
$\mu$m$^2$ in contrast shows only weak modulation. In conjunction with
these observations they simulated the emission spectrum for conditions
approximating those in the experiment apart from their choice of
thickness of the simulated plasma which was only one wavelength 
across.
In addition they used a pulse with a sine-squared temporal profile.
The simulated spectra showed a clear modulation with an 
amplitude that increased with laser intensity. In contrast to previous
work using the same code where modulations were attributed to internal
reflections of transmitted light within the plasma box (Lichters {\it
et al.\ }[6]), Watts {\it et al.} suggest that the modulation 
structure is intrinsic to harmonic generation. For a plasma one
wavelength thick and with density $n_e=13\, n_c$ they record 
modulations
at between 4 and 5 harmonics for $I \lambda_L^2 \sim 10^{19}$ W 
cm$^{-2}$ $\mu$m$^2$ across a spectral range of $m=40$ whereas for 
$I \lambda_L^2 \sim 10^{18}$ W cm$^{-2}$ $\mu$m$^2$ this range 
contracts to about $m=7$, too limited
to discern modulation even if it were present. These authors undertook
a correlation of the modulated spectra with an examination of the
dynamics of the critical surface, showing that surface modes excited by
nonlinearities could have relatively large amplitudes. For the higher
intensity studied, it appeared that harmonics at $m=4,5$ were dominant.
From this they concluded that the oscillating-mirror model of harmonic
generation proposed by Bulanov {\it et al.\ }[7] was capable of
interpreting the modulation observed.

\

\noindent The spectra reported by Watts {\it et al.\ }and the 
inferences drawn by them prompted us to look again at spectra from our
earlier simulations. The PIC code we use, like the Pfund-Lichters code
used by Watts {\it et al.}, embeds the Bourdier technique to allow for
oblique incidence but in contrast to theirs, our simulation plasma can
be up to ten wavelengths in extent, with the plasma density ramped at
the front surface.
Vacuum gaps extend from both the front of the ramp and the planar rear
surface of the plasma to the walls of the simulation box. The initial
electron temperature was chosen to be 1 keV and we used a gaussian 
pulse of variable length. The normalized quiver velocity
$a_0 \sim 0.85 {\left( I_{18} \lambda_L^2 \right)}^{1/2}$ lay in the
range 0.5-2.0 with slab densities between 4 $n_c$ and 64 $n_c$. The
ions formed a neutralizing background.

\

\noindent Our attention focused on the effect of the plasma density on
the modulated spectra. Given the wealth of detail in the spectra
presented in terms of the dependence of the modulation on laser
light intensity and pulse length as well as on the density and the 
density profile, it is perhaps worth highlighting at the start the 
principal finding from our work. 

\renewcommand{\thefigure}{3}
\begin{figure}[h]  
\centering   \leavevmode
{\bf \epsfxsize=13.0cm \epsfysize=7.0cm \epsffile{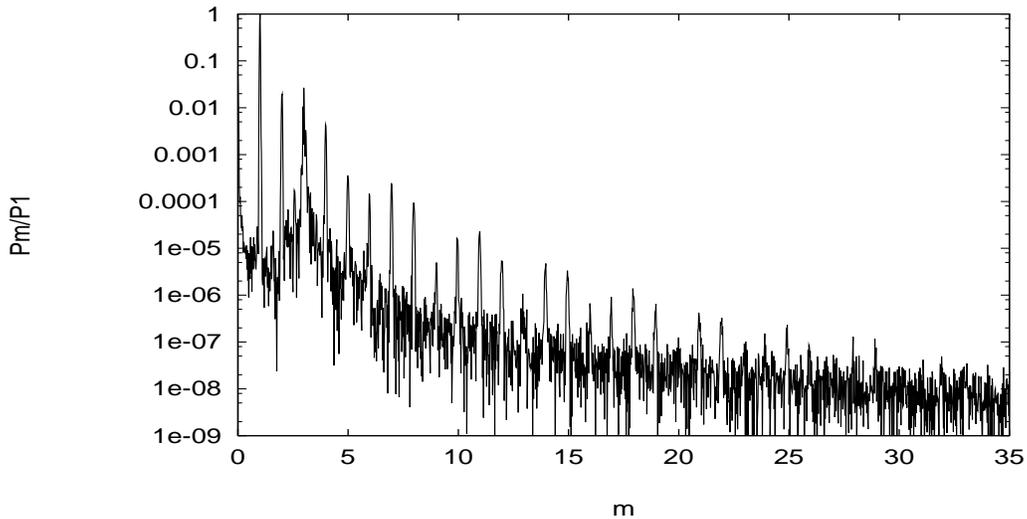} }
\caption{\small \it Harmonic modulation for $a_0=0.5, n_e/n_c=10$
and $\tau_p=17$ fs.}
\end{figure}

\noindent This is evident from Fig.\ 3 in which $a_0=0.5,\, n_e/n_c=10$
and the pulse length $\tau_p=17$ fs. The spectrum shows clear and 
uniform modulation with an average modulation frequency of 3.3
$\omega_L$ for a plasma with $\omega_p=3.2\, \omega_L$. We 
contend that the spectrum of reflected light shows harmonic structure 
modulated at the Langmuir frequency.

\

\noindent To support this interpretation we carried out simulations
across a range of intensities and pulse lengths. In all of these we 
used a
plasma slab 4 wavelengths thick without any density ramping. Fig.\ 4 
shows spectra for runs with $a_0=0.6$ and $n_e=9\, n_c$ and 19 $n_c$
respectively.

\renewcommand{\thefigure}{4}
\begin{figure}[h]  
\centering   \leavevmode
{\bf \epsfxsize=7.3cm \epsfysize=7.0cm \epsffile{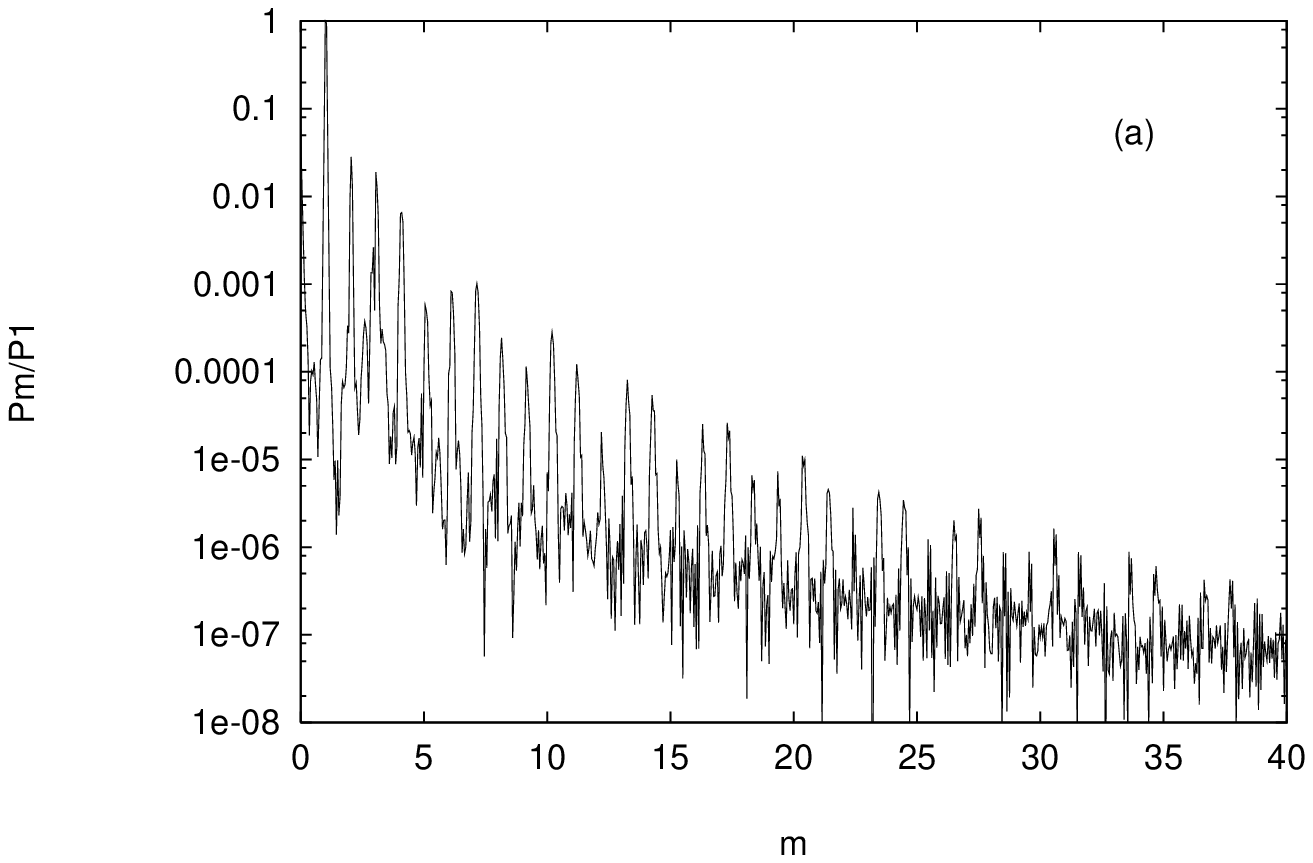} }
\hspace*{0.1cm}
{\bf \epsfxsize=7.3cm \epsfysize=7.0cm \epsffile{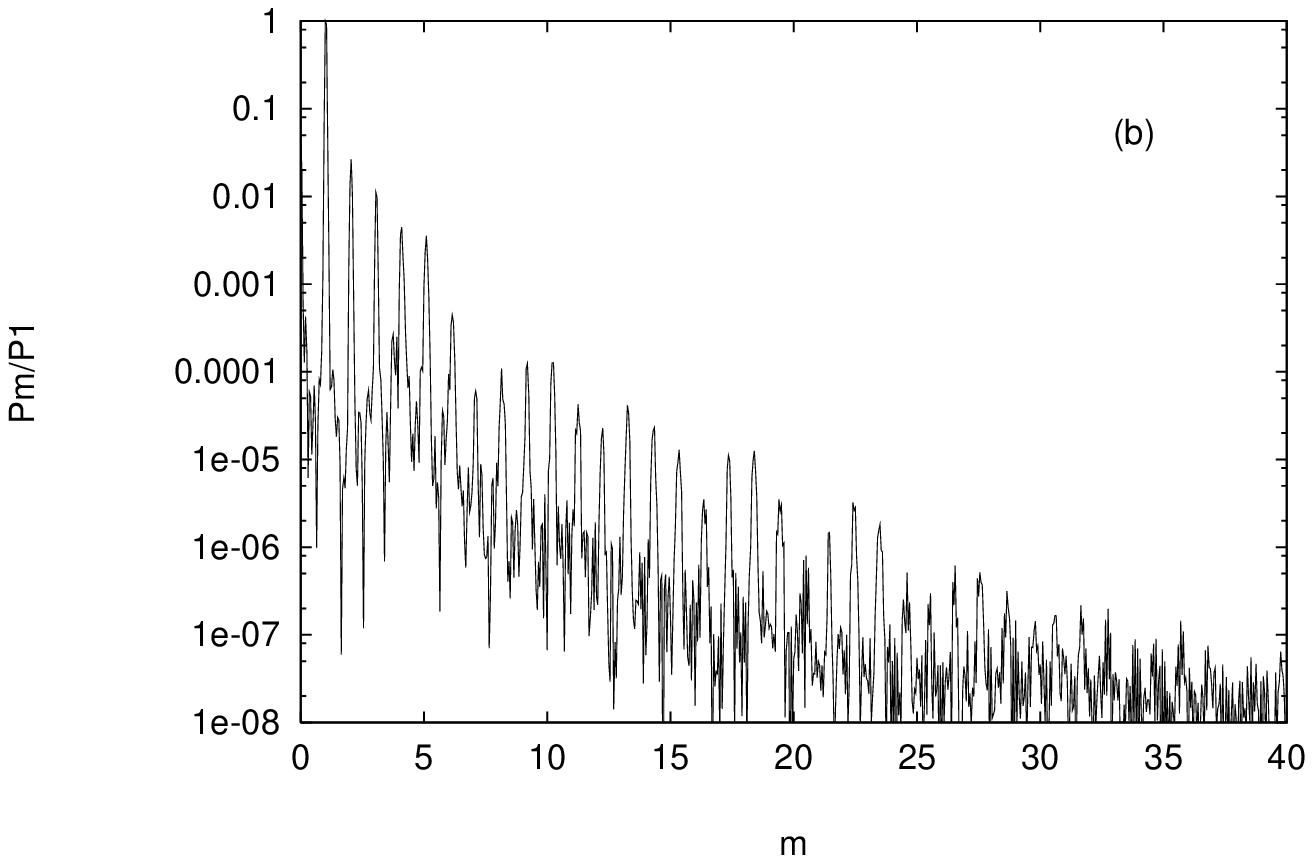} }
\caption{\small \it Harmonic modulation for $a_0=0.6, \tau_p=17$ fs,
(a) $n_e/n_c=9$ and (b) $n_e/n_c=19$.}
\end{figure}

\noindent These spectra are modulated at frequencies of 3.3 $\omega_L$
and 4.4 $\omega_L$, consistent with the Langmuir frequencies
for these densities. Note in
passing that for this combination of parameters the level of plasma
line emission is low enough for the harmonic spectrum to be relatively
unperturbed. With increasing density the harmonic range narrows
correspondingly until a point where just one modulation cycle persists
as in Fig.\ 5 for $a_0=0.5, n_e/n_c=64$ with a pulse length of 17 fs.
For this density the plasma line is centred on $m=8$ and we see the
effects of the Langmuir wave spectrum on neighbouring harmonic lines,
notably on the blue wing. Apart from the strength of the plasma line, 
this spectrum shows
somewhat similar structure to that in Fig.\ 1. It now appears that the
detail in the spectrum identified by Boyd and Ondarza [5] as the
``combination line'' is nothing other than the amplification of 
individual harmonic lines in the spectrum by the nonthermal Langmuir
waves excited in the supra-dense plasma.

\renewcommand{\thefigure}{5}
\begin{figure}[h]  
\centering   \leavevmode
{\bf \epsfxsize=7.3cm \epsfysize=6.5cm \epsffile{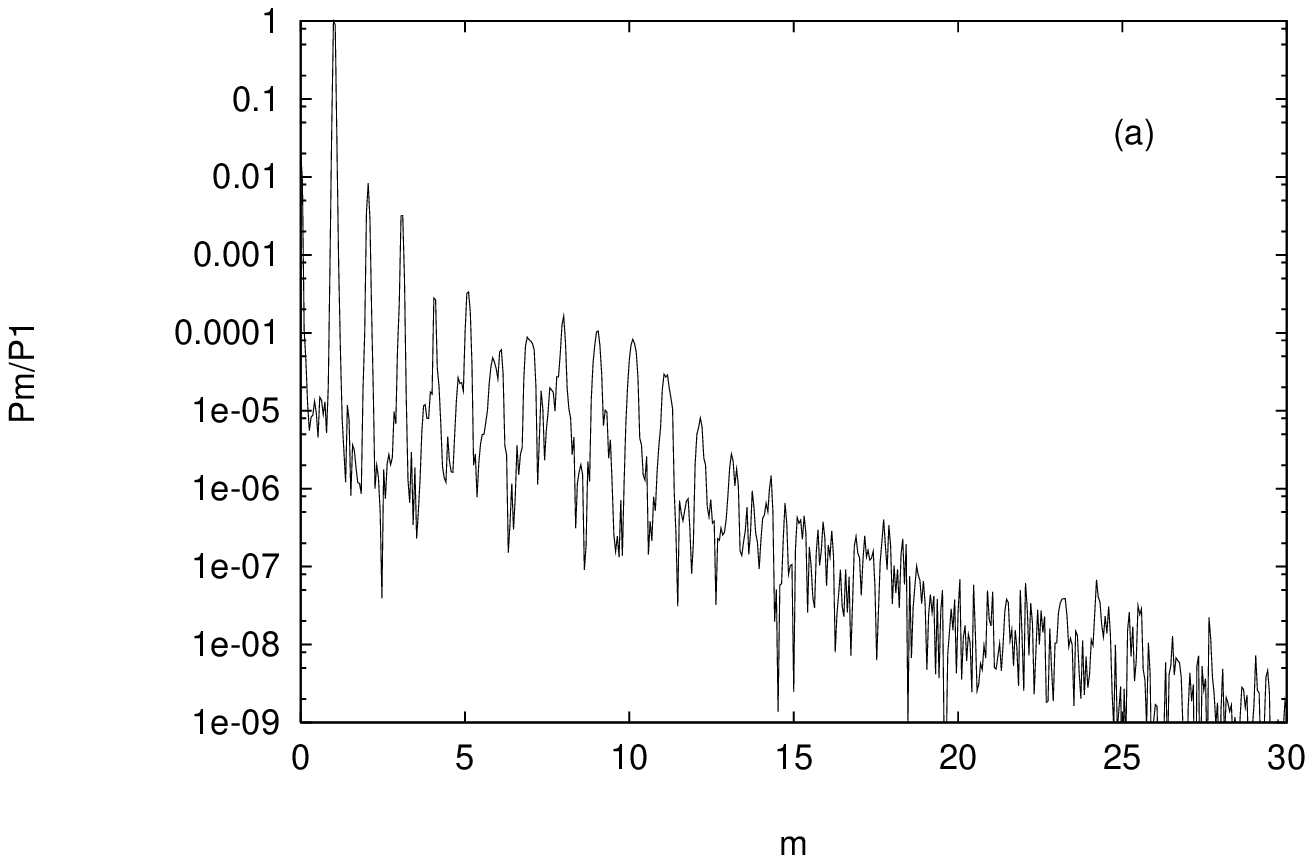} }
\hspace*{0.1cm}
{\bf \epsfxsize=7.3cm \epsfysize=6.5cm \epsffile{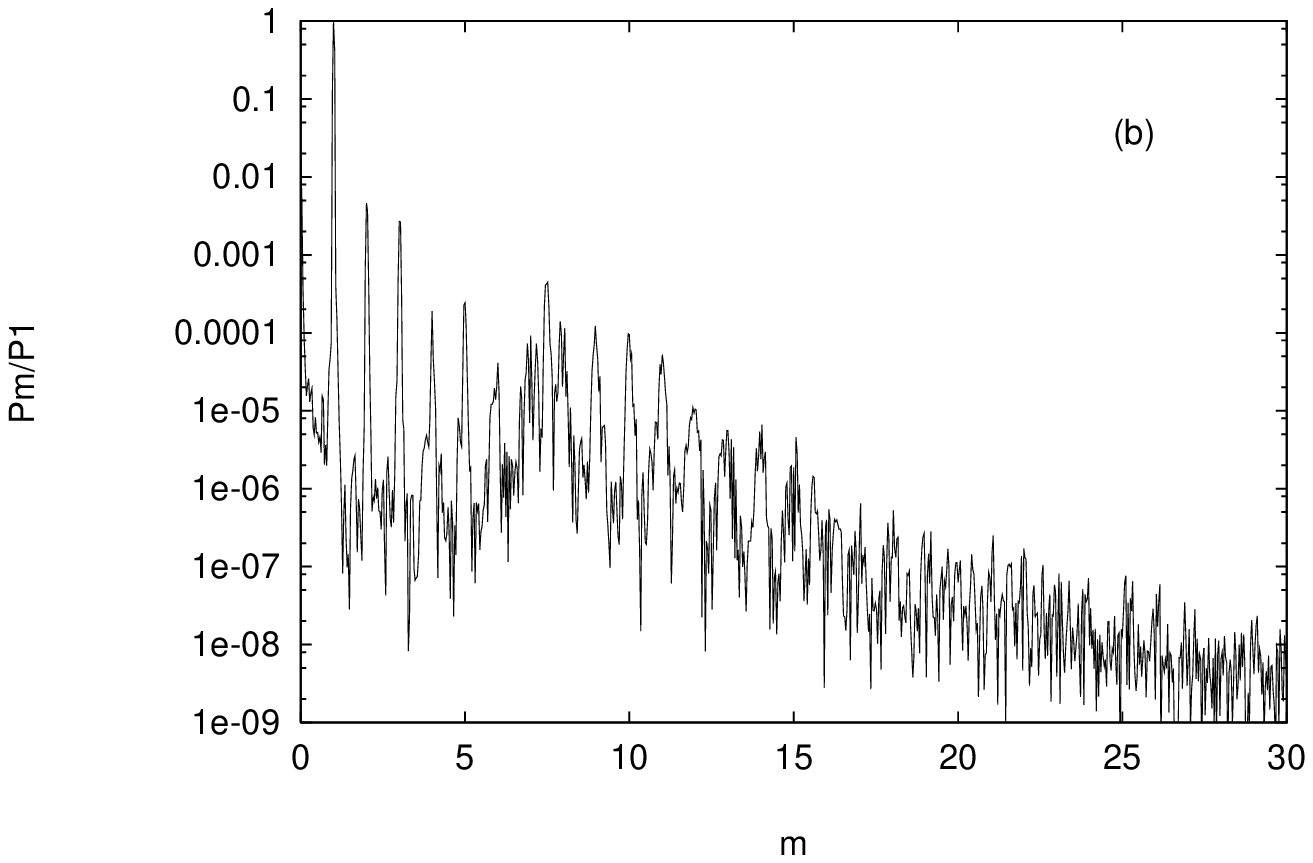} }
\caption{\small \it Harmonic modulation for $a_0=0.5, n_e/n_c=64$,
(a) $\tau_p= 17$ fs and (b) $\tau_p= 25$ fs.}
\end{figure}

\

\noindent If we wish to reproduce spectra showing several cycles of 
modulation at higher densities we need to generate emission spectra at
higher $a_0$. However the combination of higher intensity and 
increased plasma density results in the Langmuir wave being more
strongly driven and at the same time significantly broadened. Simple
linear estimates of the bandwidth of Langmuir waves $\Delta \omega$,
excited by a beam of fast electrons give $\Delta \omega \sim
\left( \Delta V_b/v_b \right){\left( n_e/n_c \right)}^{1/2}\, 
\omega_L$ for a beam velocity $v_b$ with velocity spread $\Delta V_b$.
Our PIC simulations show strongly driven Langmuir waves at the front
edge of the plasma slab and this localization is crucial for
coupling to the radiation field by the steep gradients in the
perturbed peak density zone (cf.\ Boyd and Ondarza [5]). 
As the plasma emission line becomes more strongly driven it modifies
the harmonic spectrum in the region of the plasma frequency (cf.\
Fig.\ 1). This in turn means that one has to look for modulation
cycles at frequencies well above $\omega_p$.
As we have remarked already the number of harmonic lines (above a
conversion efficiency of $ 10^{-6}$, say) reduces with increasing 
density so that the spectral window available is squeezed from both 
ends.

\

\noindent Fig.\ 6 shows a more strongly driven result ($a_0=1.0$) for
densities $n_e=25\, n_c$ and $n_e=50\, n_c$. The plasma line is now
evident and affects neighbouring harmonic lines 
so that in Fig.\ 6a with $\omega_p/\omega_L=5$ the first full
modulation cycle appears at $m=8$. Three cycles with an average  
modulational frequency of 5.3 $\omega_L$ can be resolved 
above a conversion
level of $10^{-6}$. For $n_e=50\, n_c$, $\omega_p/\omega_L \simeq 7.1$
and there are now just two cycles above this level with an average
modulational frequency of 8 $\omega_L$.

\renewcommand{\thefigure}{6}
\begin{figure}[h]  
\centering   \leavevmode
{\bf \epsfxsize=7.3cm \epsfysize=6.5cm \epsffile{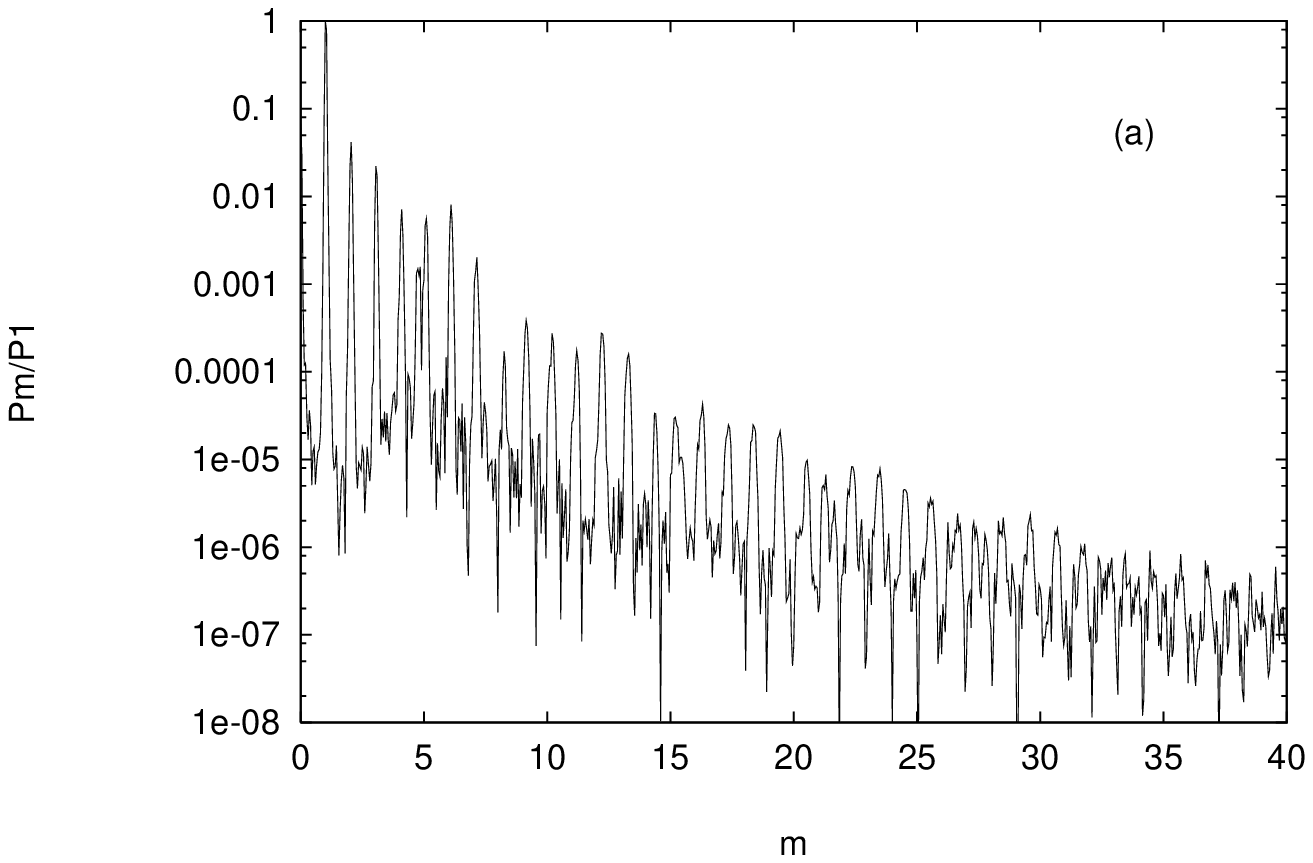} }
\hspace*{0.1cm}
{\bf \epsfxsize=7.3cm \epsfysize=6.5cm \epsffile{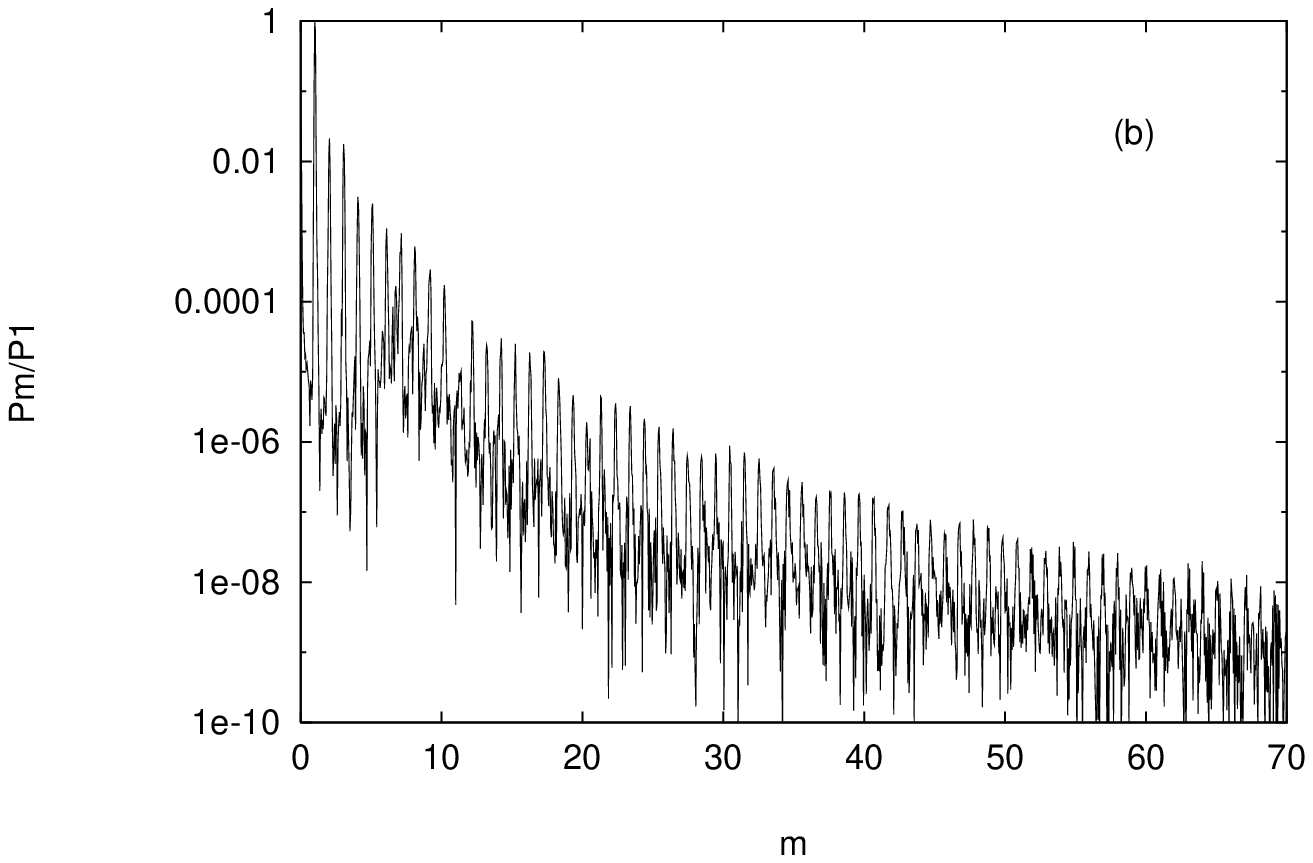} }
\caption{\small \it Harmonic modulation for $a_0=1.0, \tau_p=17$ fs,
(a) $n_e/n_c=25$ and (b) $n_e/n_c=50$.}
\end{figure}

\noindent At yet higher intensities ($a_0=2.0$) and densities
$n_e=64\, n_c$, plasma line emission is broader again than in Fig.\ 6.
The harmonic spectrum above a conversion level $10^{-6}$ now extends
to approximately $m=60$ (Fig.\ 7). However at this intensity not only is
in the plasma line a strong feature but its second harmonic is strong 
enough to perturb the line spectrum. Consequently one can resolve only
3 modulation cycles with an average frequency of 8.7 $\omega_L$.

\renewcommand{\thefigure}{7}
\begin{figure}[h]  
\centering   \leavevmode
{\bf \epsfxsize=13.0cm \epsfysize=6.6cm \epsffile{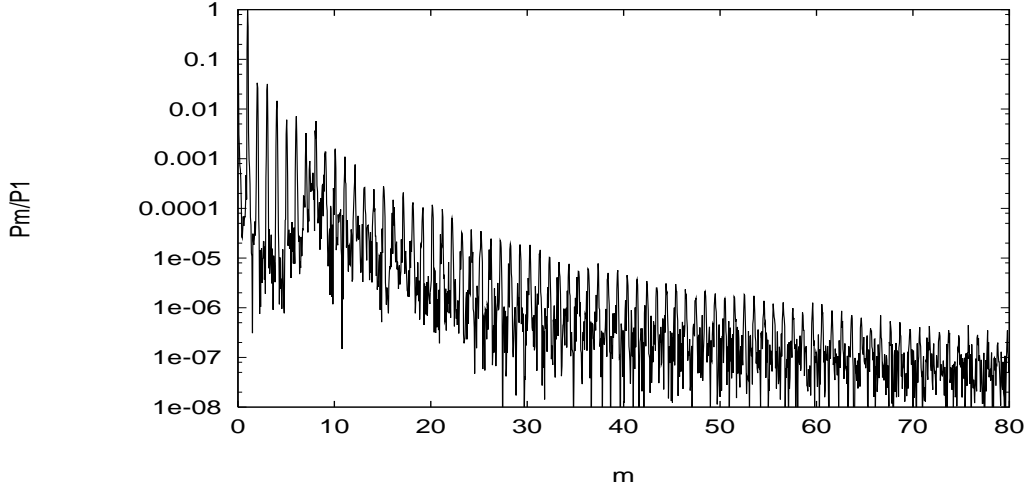} }
\caption{\small \it Harmonic modulation for $a_0=2.0, \tau_p=17$ fs,
and $n_e/n_c=64$.}
\end{figure}

\noindent In all the cases discussed thus far, the PIC simulation
plasma has been in the form of a slab. We have also examined the
effect on the modulation of a ramped density profile to check the claim
by Watts {\it et al.\ }that density ramping served to suppress the 
effect. We too found a comparable decay in modulation as the thickness
of the ramp is increased. Fig.\ 8 shows our result for $a_0=0.5$,
$n_e/n_c=10$ for different thickness of ramp $\Delta$;
(a) $\Delta=0.2\, \lambda_L$ and (b) $\Delta=0.4\, 
\lambda_L$. From this it is evident that although some modulation
persists when a ramp of thickness $0.2\, \lambda_L$ is present,
the effect disappeared completely in Fig.\ 8b.

\renewcommand{\thefigure}{8}
\begin{figure}[h]  
\centering   \leavevmode
{\bf \epsfxsize=7.0cm \epsfysize=6.6cm \epsffile{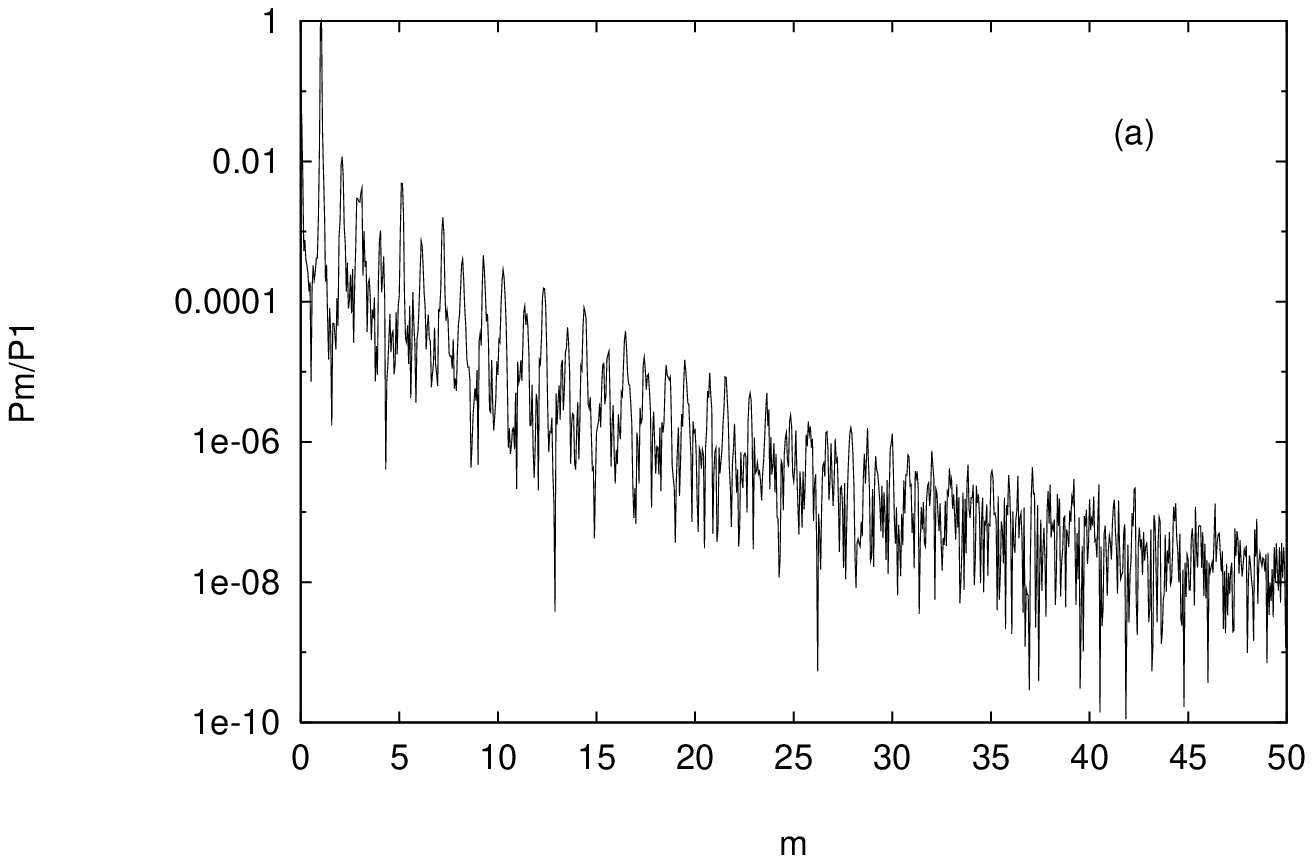} }
\hspace*{0.1cm}
{\bf \epsfxsize=7.0cm \epsfysize=6.6cm \epsffile{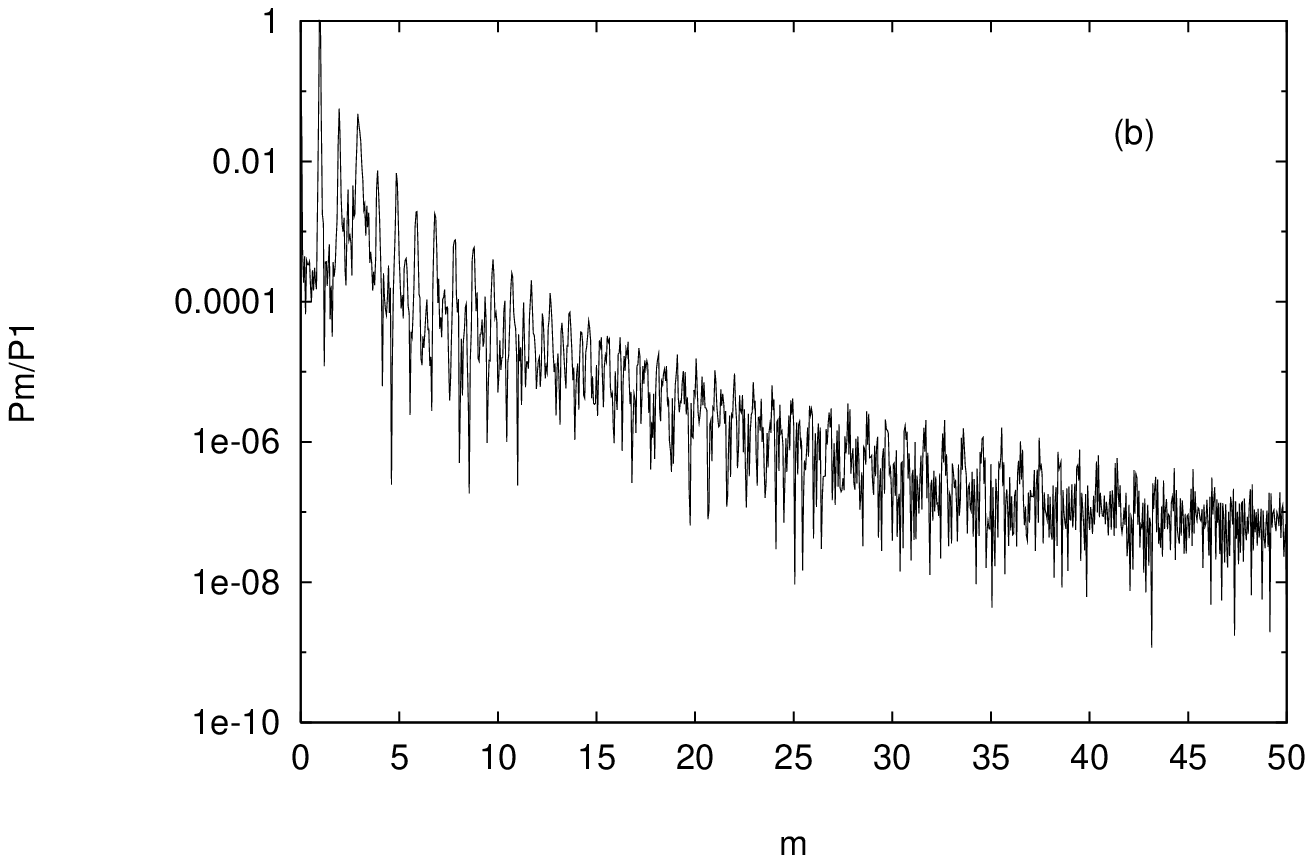} }
\caption{\small \it Harmonic modulation for $a_0=0.5, \tau_p=17$ fs,
 $n_e/n_c=10$, (a) $\Delta=0.2\, \lambda_L$ and (b) $\Delta=0.4\, 
\lambda_L$.}
\end{figure}

\newpage

\begin{center}
\renewcommand{\thesection}{II.}
\section {\large \bf CONCLUSIONS}
\end{center}

\vspace*{0.5cm}

\noindent Our simulations lead us to conclude that the modulations
we see in the harmonic spectrum stem from the effect of Langmuir
waves excited in the plasma slab. We have shown that the modulation
frequency varies with the plasma density as ${n_e}^{1/2}$ over a wide 
range. The modulation is evident for choices of intensity and pulse
length for which plasma line emission does not distort the spectrum
to any marked degree. On this interpretation the modulation
frequency affords a measure of plasma density as distinct from the
claim by Watts {\it et al.\ }of its potential as a diagnostic of
the dynamics of the critical surface. The presence of a density ramp
a fraction of a wavelength thick appears to suppress the modulation
as Watts {\it et al.\ }found. We see this as consistent with our
interpretation in that the presence of a ramp ensures that there is 
no longer a unique plasma density at the surface.

\

\begin{center}
\noindent {\large \bf Acknowledgments}
\end{center}

\vspace*{1mm}

\noindent One of us (ROR) acknowledges support from CONACyT under 
Contract No.\ 43621-F.

\

\small

\begin{center}
\noindent {\large \bf REFERENCES}
\end{center}
\vspace*{1mm}

\noindent [1] Watts I.\ {\it et al., Phys.\ Rev.\ Lett.\ }{\bf 88} 
          (15), 155001 (2002).

\noindent [2] Norreys P.\ A. {\it et al., Phys.\ Rev.\ Lett.\ }
           {\bf 76}, 1832 (1996).

\noindent [3] Gibbon P., {\it Phys.\ Rev.\ Lett.\ }{\bf 76}, 
          50 (1996).

\noindent [4] Boyd T.J.M.\ and Ondarza-Rovira R., in {\it
          Proceedings, International Conference on Plasma Physics},
          \hspace*{5mm}Nagoya, September 1996, edited by H.\ Sugai and 
          T.\ Hayashi (Japan Society of Plasma Science in 
          \hspace*{5mm}Nuclear Fusion Research, Nagoya, 1997), 
          Vol.\ 2, p.\ 1718.

\noindent [5] Boyd T.J.M.\ and Ondarza-Rovira R., {\it Phys.\ Rev.\ 
          Lett.\ }{\bf 85} (7), 1440 (2000).

\noindent [6]  Lichters R. {\it et al., Phys.\ Plasmas} {\bf 3}, 
          3425 (1996).

\noindent [7] Bulanov S.\ V.\ and Naumova N.\ M., {\it Phys.\
          Plasmas} {\bf 1}, 745 (1994).
 
\end{document}